
\def\spose#1{\hbox to 0pt{#1\hss}}
\def\lta{\mathrel{\spose{\lower 3pt\hbox{$\mathchar"218$}}
     \raise 2.0pt\hbox{$\mathchar"13C$}}}
\def\gta{\mathrel{\spose{\lower 3pt\hbox{$\mathchar"218$}}
     \raise 2.0pt\hbox{$\mathchar"13E$}}}
\magnification=\magstep1
\hsize 6.5 truein
\vsize 8.6 truein
\parindent 20pt
\centerline{INTERSTELLAR GRAINS IN ELLIPTICAL GALAXIES: GRAIN EVOLUTION$^1$}
\vskip .5in
\centerline{John C. Tsai$^{2,4}$ and William G. Mathews$^3$}

\vskip .2in
\centerline{$^2$NASA Ames Research Center}
\centerline{Mail Stop 245--3}
\centerline{Moffett Field, CA 94035--1000}

\vskip .2in
\centerline{$^3$University of California Observatories/Lick Observatory}
\centerline{Board of Studies in Astronomy and Astrophysics}
\centerline{University of California, Santa Cruz, CA 95064}

\vskip .2in
\centerline{$^4$Canadian Institute for Theoretical Astrophysics}
\centerline{McLennan Labs, University of Toronto}
\centerline{Toronto, ON, Canada M5S 1A7}
\vskip 1.8truein

\noindent
Received:

\noindent
PROOFS TO BE SENT TO:

\noindent
----------------------

\noindent
\+ Lick Observatory\cr
\+Santa Cruz, CA 95064\cr

\+Subject headings:\cr
\noindent
$^1$UCO/Lick Observatory Bulletin No.

\par\vfill\eject
%
%
%
\centerline{ABSTRACT}
We consider the lifecycle of dust introduced into the hot interstellar
medium in isolated elliptical galaxies.  Dust grains are ejected into
galactic-scale cooling flows in large ellipticals by normal mass loss
from evolving red giants.  Newly introduced dust rapidly enters the hot
interstellar plasma and is sputtered away by thermal collisions with ions
during the slow migration toward the galactic center in the cooling flow.
Before the grains are completely sputtered away, however, they emit
prodigious amounts of infrared radiation which may contribute to the
large far infrared luminosities observed in ellipticals.  The infrared
emission depends critically on the sputtering rate.  Since our
understanding of both the plasma and radiation environments in ellipticals
is quite good, these galaxies provide an excellent venue for studying
the physical processes of dust grains and perhaps also their composition
and size distribution.

In order to study the global properties of grains in ellipticals
we construct a new series of King-type galactic models which are
consistent with the fundamental plane, galactic mass to light ratios and
other relevant observational correlations.  We describe a new
``continuity'' procedure to construct simple time-dependent gas dynamic
models for cooling flows.

Although grains can flow a considerable distance from their radius of
origin in the hot interstellar medium of some galaxies before being
sputtered away, we show that the grain size distribution at every radius
is accurately determined by assuming {\it in situ} sputtering of dust
grains, completely ignoring advection.  This occurs since the stellar
density profile is so steep that the majority of grains at any
galactic radius is produced locally.

Although thermal sputtering destroys the grains, we show that the
dominant source of grain heating is absorption of starlight; grain heating
by collisions with energetic thermal electrons or X-ray absorption are
negligible.  Previous studies have claimed that the loss of thermal energy
from a hot, dusty plasma is dominated by grain heating via electron-grain
collisions and subsequent IR radiation.  However, we show that when
self-consistent grain sputtering is included the dust-to-gas ratio is
reduced and radiative cooling, not electron-grain interactions dominates
plasma cooling, even for the most massive ellipticals.  This conclusion
is insensitive to the grain size distribution assumed for the stellar ejecta.
\vskip .2in
\centerline{1. INTRODUCTION}

Elliptical galaxies, long thought to be simple ensembles of
non--interacting stars, are now known to be complicated
multicomponent systems of stars and gas.
The interstellar medium (ISM) in ellipticals is
dominated by hot gas which is slowly cooling by
radiative losses; as a result the ISM gradually flows inward
as a cooling flow.
X--ray observations of optically bright ellipticals reveal
gas masses of
$\sim 10^8 - 10^9\rm M_\odot$ at high temperatures
$\sim 10^7\rm K$ (Forman et al. 1979; Nulsen, Stewart, \& Fabian
1984; Forman, Jones, \& Tucker 1985; Canizares, Fabbiano, \& Trinchieri
1987).
Radio and optical observations (e.g., 21 cm, $\rm H\alpha$, and CO)
indicate that elliptical galaxies typically
contain $\sim 10^7 - 10^8 \rm M_
\odot$ of cold gas usually located in the core regions
(see e.g. Kormendy \& Djorgovski
1989 for a review).

In view of the high temperature of the ISM it is remarkable that
elliptical galaxies also contain substantial amounts of dust.
Optical images of $\sim 50$\% of all ellipticals exhibit dust
lanes, disks, or patches (e.g., Sadler \& Gerhard 1985;
Kormendy \& Stauffer 1987; Ebneter et al. 1988).
Since dusty
disks would be difficult to see when viewed face on,
it is possible that all
ellipticals contain appreciable dust.
Further evidence for dust are the large
far-infrared IRAS luminosities detected in
approximately 50\% of ellipticals.
The ratio of flux energy $\nu F_{\nu}$ at 100$\mu {\rm m}$ to optical
B--band emission has a cosmic spread,
$\nu_{100}F(100\mu{\rm m})/\nu_B F(\nu_B) = 0.006 - 0.088$
(Jura et al. 1987);
for cD galaxies this ratio is about ten times larger
(Bregman, McNamara, \& O'Connell 1990).

The geometrical distribution of dust and the associated infrared
emission in ellipticals may be either concentrated near the galactic core
or distributed throughout the stellar system.  Concentrated patches
or disks of dusty gas observed near the galactic cores could arise
either (i) from merger events with galaxies having large amounts of
cool, dusty gas and/or (ii) from cold gas deposited by galactic cooling flows
as the hot ISM slowly radiates away its thermal energy.
These two possibilities may be distinguished observationally.
Because of the destruction of grains in the hot ISM by
sputtering, the dust to gas ratio is expected to be much
lower, at least initially, in cool gas condensed from cooling flows.
Moreover, gas deposited by galactic
cooling flows should exhibit the same global rotational dynamics
and sense of rotation as that of the stellar system
(Kley \& Mathews 1995).

In addition to the central dust patches,
more widely distributed dust is continuously generated
by stellar mass loss from evolving red giants throughout the galaxy.
Knapp, Gunn, \& Wynn--Williams (1992) have shown that the
relatively hot dust that emits
12$\mu$m radiation in ellipticals is spatially distributed like the galactic
stars and may therefore be circumstellar.
However, because of the dynamical interaction of the
dusty stellar ejecta with the
hot ISM, the time that recently ejected gas and dust lingers
near the parent star is limited.
When the dusty stellar ejecta interacts with the
hot $T \sim 10^7$K ISM, numerous hydrodynamic instabilities are expected
(Mathews 1990) resulting in breakup and deceleration by drag
forces as the ejecta trails off behind the star.
The fragmented new gas is rapidly heated by thermal conduction as it
merges with the hot ISM, even in the presence of a magnetic field.
It is easy to show (Mathews 1990)
that the cool stellar ejecta is separated from its
parent star and melts into the ISM on
time scales $\sim 10^4 - 10^5$ yrs, so fast that
little or no grain destruction can occur until the dust is immersed
in the hot ISM.
Once in contact with the hot ISM of number density $n$,
grains with radius $a$ are sputtered away
by thermal collisions with protons and helium nuclei; the
sputtering lifetime is
$t_{sp} \approx 10^9(a/\mu {\rm m})(n/10^{-3}{\rm cm}^{-3})^{-1}$
yrs provided ISM temperatures are $T > 2 \times 10^{6}$ K
(Tielens et al. 1994; Draine \& Salpeter 1979; Seab 1987).
However, during their short lifetimes as they sputter away
and are advected inward by the galactic cooling flow,
the grains can nevertheless
emit a huge amount of far IR radiation
(Dwek 1986; 1987).
But the dust-to-gas ratio and IR emissivity from heated grains
is greatly reduced by the short sputtering lifetimes
so emission and sputtering must be treated in a self-consistent
manner.
Emission from this distributed dust may in fact contribute
substantially to the observed IRAS luminosities.
Unfortunately the spatial resolution of IRAS observations
($3'\times 5'$ at 100$\mu$m) is insufficient
to discern whether the emission comes from the dusty clouds
near the galactic centers or is distributed throughout
the galaxies.

In this and a subsequent paper we study the physical properties
and emission from the distributed dust component.
The infrared emission from distributed dust associated
with stellar mass loss must be present in every elliptical
and is a non-trivial lower limit to the observed fluxes.
Fortunately the dust environment --
particularly the gas density
and temperature in the ISM -- is well known from
successful hydrodynamic models of cooling flows
(e.g. Loewenstein \& Mathews 1987).
The local gas density depends on
the rate of mass loss from an aging stellar population
$\alpha_*(t)\rho_*(r)$ g cm$^{-3}$ s$^{-1}$
which is well known and insensitive to the
stellar IMF (Mathews 1989).
While grains are destroyed by ion impacts from the hot
plasma, we find that they are heated primarily by
the local radiation field.
The radiation within ellipticals
is also very well known in both the optical-IR
(Renzini \& Buzzoni 1986; Bruzual 1985)
and in the near UV as observed with the Hopkins UV
Telescope (e.g. Ferguson \& Davidsen 1993).
Soft X-ray and far--UV radiation from the hot ISM can be
determined either theoretically from an atomic emissivity code
(Raymond 1991) or from direct observation
(e.g. Trinchieri, Fabbiano,
\& Canizares 1986; Serlemitsos et al. 1993).
{\it Our excellent knowledge of the radiation and plasma
environment makes the interstellar medium in ellipticals
an excellent laboratory for studying the physics of
grain sputtering and emission.}
In comparing theoretical and observed
infrared emission the primary uncertainties are the
initial distribution of dust grain sizes
and the grain composition in stellar ejecta;
indeed, information about these parameters can be
found by detailed comparison of theoretical models
with infrared observations.
A detailed study of the expected IR spectrum from ellipticals
and their emission into each
IRAS band will be the subject of a subsequent paper.

In this paper we discuss the life cycle,
thermal properties, and size distribution of
dust grains ejected into
elliptical galaxy cooling flows (ISM) by evolving stars.
In order to predict the infrared luminosity due to
dust emission from a galaxy
of given $L_B$ it is necessary to know the sputtering
rate, the grain heating and cooling rates,
the temperature and infrared emissivity
of grains of all sizes and at all galactic radii.
Since we want to explore the variation of
infrared to optical luminosities among ellipticals of various
masses, we develop in \S2 a scheme for generating
models of the radial stellar distribution
within elliptical galaxies that is consistent with the fundamental
plane and other observed galactic properties.
In \S3 we describe a new procedure to develop simple
time-dependent models for the cooling hydrodynamics
in the model galaxies.
These cooling flows are used to study the effects of sputtering
and advection on grain size distributions in \S4.
In \S5 we compare various grain heating
mechanisms: by ambient radiation,
by electron and ion impacts,
and by X-ray absorption.
We conclude that grain heating by starlight dominates
all other heating sources.
Grain size distributions for each model galaxy
are presented in \S6.
In \S7 we consider the effects of the presence
of grains on the temperature of the ISM;
we show that cooling of the hot ISM
by thermal collisions with grains is generally small
compared to optically thin radiative losses.

\vskip .2in
\centerline{2. SELF-CONSISTENT MODELS FOR ELLIPTICAL GALAXIES}

For analytic simplicity and for comparison with previous theoretical
studies we assume spherical, King-type model galaxies having
stellar density profiles given by
$$ \rho_*(\xi) = \rho_{*o} ( 1 + \xi^2)^{-{3\over 2}}; ~~~~ \xi \equiv
r/r_c \eqno(1)$$
where $r_c$ is the stellar core radius.  The stellar distribution is
terminated at an outer boundary $r_t$, or
$\xi_t = r_t/r_c$.  From eq. (1), the stellar mass within radius
$\xi$ in units of the stellar core mass $M_{*c} = 4 \pi r_c^3 \rho_{*o}/3$
is
$$\mu_{mass}(\xi) = 3\left[ \ln [\xi + (1 + \xi^2)^{1\over 2}]
- \xi (1 + \xi^2)^{-{1\over 2}}\right] \eqno(2)$$
and the mass surface density
$$\Sigma_*(\lambda) = {2 \rho_{*o} r_c
\over (1 + \lambda^2)} \left(1 - { 1 + \lambda^2 \over 1 +
\xi_t^2} \right)^{1\over 2} \eqno(3)$$
depends on the projected radius
$\lambda = R/r_c$.

The parameters that determine the form of the stellar distribution within
each galaxy -- $\rho_{*o}$, $r_c$ and $r_t$ -- must be consistent with
observational constraints on the properties of ellipticals,
particularly the fundamental plane relation (e.g. Dressler et al.
1987; Djorgovski \& Davis 1987).  In Appendix A
we describe the procedure
we have used to derive the stellar properties of a
one parameter family of galaxy models.
Relevant properties of the three
representative galaxies used in our study of dust evolution
are listed in Table 1.  Note that the listed ellipticals extend
from the brightest only to those of intermediate luminosity
($L_B \lta 10^{10}$ $L_\odot$) for which the presence cooling flows
can be reasonably assured.  The global X-ray
emission from ellipticals of lower luminosity is dominated by
stellar sources, evidently low mass X-ray binaries.
In addition, low luminosity ellipticals
have shallower potential wells and the ISM can
become a galactic wind if the collective luminosity
of Type Ia supernovae exceeds the rate of emission from the hot ISM.
We assume that most or all moderate to massive ellipticals
have cooling flows as evidenced by their large X-ray
luminosities $L_x$.  Recent observations of low iron abundances
in several bright ellipticals (Serlemitsos et al. 1993; Forman et al. 1993;
Mushotzky et al. 1995) indicate that the Type Ia supernova rate is
sufficiently low to permit cooling flows in all galaxies considered
here (Loewenstein \& Mathews 1987; 1991).

In order to estimate the sputtering rate of grains in the ISM,
it is necessary to have estimates of
the gas temperature, density and velocity in the cooling flows.
We describe how the gas density and velocity are
determined in \S 3.  The temperature,
ideally, should be determined by observations of the X--ray emitting gas
in galaxies having a spread in luminosities.  However, current temperature
determinations are available for only a few galaxies (e.g., NGC 4472 and
NGC 1399: Serlemitsos et al. 1993; NGC 4636: Mushotzky et al. 1995)
and are restricted to the highest
luminosity cases.  We therefore adopt a theoretical approach for determining
the temperatures.  Gas ejected by stars in the galaxy is rapidly virialized
to the mean stellar temperature $\langle T_* \rangle$ which characterizes
the average stellar velocity dispersion
(weighted by the stellar mass density) as stars orbit in the total galactic
potential.  Assuming most of the interstellar gas comes from stars and that
other sources of heating are small, the temperature of the gas should be
very close to $\langle T_* \rangle$.
This has been confirmed by detailed hydrodynamic calculations.
To estimate $\langle T_* \rangle$ we
adopt appropriate density structures for the galactic dark matter halos
and solve the equation of stellar hydrodynamics.  This is described in
Appendix A.  The temperatures listed in Table 1 are computed in this manner.

As a check of the above procedure, we plot in Figure 1 the
($L_B$,$\langle T_*\rangle$)-relation implied by our procedure for
determining the gas temperature.  The location of several of our model
galaxies are indicated by filled circles and the general relation is
given by the solid line.  Three galaxies with recent gas temperature
determinations by {\it ASCA}, NGC 4472, NGC 4406, and NGC 4636
(Makishima 1994; Mushotzky et. al. 1995) are also plotted.  The observed
temperatures of these galaxies agree reasonably well with our predicted
($L_B$,$\langle T_*\rangle$)-relation.  A more accurate determination of
the temperatures is not required for the considerations of this paper.
\vskip .2in
\centerline{3. TIME-DEPENDENT ``CONTINUITY'' MODELS FOR GALACTIC COOLING FLOWS}

For three nearby elliptical galaxies which were well resolved
by the {\it Einstein} Observatory Imaging Proportional Counter (IPC),
Trinchieri, Fabbiano, \& Canizares (1986) found
that the X-ray surface brightness distribution is very similar to that
of the starlight, i.e.  $\Sigma_x(\lambda) \propto \Sigma_*(\lambda)$.
Since the X-ray emissivity is proportional to the square of the
gas density, it follows from eq. (1) and the radial constancy
of the stellar mass to light ratio that the gas density varies as
$$ \rho(\xi) = \rho_o (1 + \xi^2)^{-{3\over 4}}.\eqno(4)$$
We have implicitly assumed that the gas temperature
is independent of galactic radius.
Isothermality throughout most of the volume of the cooling flow
is a general theoretical result (Loewenstein \& Mathews 1987; 1991)
and is supported by recent BBXRT and {\it ASCA} observations
(Serlemitsos et al. 1993; Makishima 1994; Mushotzky et al. 1995).
Within the galactic cores,
however, the gas must cool rapidly and large deviations from
isothermality are expected.
The gas temperature may also decrease at the very largest radii
since $T_*(r_t) \rightarrow 0$ if the stellar velocities become
preferentially radial at large radii.

Since the gas density distribution $\rho(\xi)$ of eq. (4) has the
same core radius and total radius as the stellar distribution,
$\rho(\xi)$ is completely specified by the central density
$\rho_o$.  This can be determined by relating the gas density to the X--ray
luminosity in the {\it Einstein} IPC band ($\sim 0.2$ to $\sim 4$ keV),
$$ \eqalign{  \left(\rho_o \over  m_p\right)
=0.165~ \left({\Lambda_
{\Delta E}\over 10^{-23}\rm ergs~cm^{3}~s^{-1}}
\right)&^{-{1\over 2}} \left( {L_x\over 10^{41}~ {\rm
ergs~s^{-1}}}\right)^{1\over 2}\cr
& \left({r_c\over {\rm kpc}}\right)
^{-{3\over 2}} \left[ {\rm ln}\left( 2\xi_t\right)
- 1\right]^{-{1\over 2}}~{\rm cm^{-3}}}~
\eqno(5)$$
where $m_p$ is the proton mass. The atomic emissivity in
the given energy band, $\Lambda_{\Delta E}$, which we have determined
from the code of Raymond (1991), depends only on the gas temperature
(although this dependence is weak for the temperatures of interest).
The X--ray luminosity is observed to correlate with
the B--band optical luminosity
$$ L_x = 10^{17.85}~ \left( {L_B\over L_\odot}
\right)^{2.18}~h^{2.36}~  {\rm ergs~s^{-1}} \eqno(6)$$
(Donnelly, Faber, \& O'Connell 1990) where $h$ is the
Hubble constant normalized to $\rm 100$ $\rm km~s^{-1}$ $\rm Mpc^{-1}$.
We assume $h=0.75$ throughout.

Having specified the gas density distribution, we need to determine the
velocity of the gas in the cooling flow $u(r)$
to complete the description of
the model galaxies.
Although the inward gas flow velocity in galactic cooling flows is highly
subsonic, these flows are not in steady state since the source term
for new gas expelled by evolving stars varies considerably during the time
required for gas to flow across the galaxy.  The specific mass loss rate
from a single-burst population of evolving
stars is $\alpha_*(t) = \alpha_n(t/t_{now})^{-p}$ where $\alpha_n \approx
1.6 \times 10^{-12} \rm yr^{-1}$, $p = 1.35$ and $t_{now}$ is the
present time.
The time variation of $\alpha_*(t)$ is insensitive to
the IMF assumed for the galaxy (Mathews 1989).

Cooling flows were not present at the very earliest times in the
history of massive ellipticals.
The significant iron abundance in clusters of galaxies suggests that
a considerable amount of enriched
mass processed in stars was ejected long
ago from the elliptical galaxies
(or their progenitors) by galactic winds driven by Type II supernovae
(Mathews 1989; David, Forman, \& Jones 1991).
It is therefore reasonable to assume that the cooling flows observed
in massive ellipticals today began at some time $t_{cf}$ in the
past when the galactic outflow subsided.

We now show that,
given several simplifying assumptions, the velocity field of
the cooling flow at the present time $u(r,t_{now})$
can be found from the continuity equation alone,
$$ {\partial\rho \over \partial t}~+~ { 1 \over r^2 } {\partial \over
\partial r}(r^2 \rho u) = \alpha_*(t)\rho_*.\eqno(7)$$
The inward flow velocity $u(r,t)$ is the only unknown dependent
variable if we note that
the stellar density $\rho_*$ does not vary significantly with time,
and require that the gas density at time $t$ vary as
$$\rho\left(\xi,t\right) = \left({t\over t_{now}}\right)^{-s}
\rho\left( \xi\right),\eqno(8)$$
where $\rho\left( \xi\right)$ is the gas density at $t_{now}$
given by eq. (4) and $s\approx p/2$.
Condition (8) is based both on the results of detailed hydrodynamical
calculations (Loewenstein \& Mathews 1987; David, Forman, \& Jones 1991)
and on the approximate considerations described in Appendix B.
During the evolution of cooling flows the
X-ray luminosity varies with the stellar mass loss rate,
$L_x\propto \alpha_*(t)$.
But $L_x$ at any time is proportional to the square of the gas
density at any radius, $L_x \sim
\Lambda_{\Delta E}~ \rho^2$.
If the gas temperature (and also $\Lambda_{\Delta E}$)
remains roughly uniform, it follows from these two relations that
$\rho \propto \alpha_*(t)^{1/2} \propto t^{-p/2}$ as in eq. (8)
(see Appendix B).

Substituting eqs. (1) and (8) into eq. (7), we get an equation
containing only a spatial derivative,
$$ -s \gamma\Delta\left(\xi\right)\theta^{-s-1}~+~{\theta^{-s}\over
\xi^2}{\partial\over \partial\xi}\left[\xi^2\Delta\left(\xi\right)
\eta\left(\xi,\theta\right)\right]~=~\theta^{-p}\Delta_*\left(\xi\right)
\eqno(9)$$
where $\theta = t/t_{now}$, $\gamma=\rho_o/\alpha_n\rho_{*o}t_{now}$ is a
dimensionless parameter, and the fluid variables are expressed in
non--dimensional form:
$$\Delta_*\left(\xi\right)= {\rho_*\left(\xi\right)\over
\rho_{*o}}=\left(1 + \xi^2\right)^{-{3\over 2}},\eqno(10)$$
$$\Delta\left(\xi\right)= {\rho\left(\xi\right)\over
\rho_{o}}=\left(1 + \xi^2\right)^{-{3\over 4}},\eqno(11)$$
and
$$\eta\left(\xi,\theta\right)={u\left(\xi,\theta\right)\over
u_o};~~~~ u_o={r_c\alpha_n\rho_{*o}\over \rho_o}.\eqno(12)$$

Eq. (9) is valid at all times after the onset of the cooling flow
($t_{cf}$).  It must therefore be valid at the present time, where
$\theta = 1$.  On integrating over $\xi$, we determine the
velocity field at the present time,
$$ \eta\left(\xi,1\right) = - {1\over \Delta\xi^2}
\int_\xi^{\xi_t}
\left[\Delta_* + s\gamma\Delta\right]~\xi'^2 d\xi'.\eqno(13)$$
where no flow across the outer boundary of the
galaxy is assumed.

Eq. (13) represents an approximation for the velocity field
in a fully time dependent galactic cooling flow.
Velocities computed from eq. (13) for the three galaxies of Table 1
are shown in Figure 2.  The smallest galaxy has the smallest velocity at
any given value of the normalized radius $\xi/\xi_t$
because the value of $\gamma$ is least for the smallest galaxy.  The
velocity profiles compare favorably with those derived from detailed
hydrodynamical calculations (e.g. Loewenstein \& Mathews 1987) as do the
flow times predicted from the velocities (see Figure 3).
\vskip .2in
\centerline{4. GRAIN SPUTTERING}

Dust grains introduced into hot gas are eroded by collisions with
energetic protons and helium ions in the gas.  The rate at which dust
particles are sputtered away has been computed in detail by Tielens et
al. (1994) and by Draine \& Salpeter (1979) for both graphite and
silicate grains.  We adopt the analytic form
$$ {da\over dt} = - h~{\left(\rho\over m_p\right)}
\left[\left({T_d\over T}\right)^\omega + 1\right]^{-1} \eqno(14)$$
for the rate at which the radius of the dust grain $a$ decreases with
time in a hot plasma of temperature $T$ and density $\rho$.
This relation is a good approximation to the detailed
calculations for both graphite and silicate when $h = 3.2\times
10^{-18} {\rm cm^4~s^{-1}}$, $\omega = 2.5$, and $T_d = 2\times 10^6~\rm K$.

Eq. (14) can be used to compute the local sputtering time for a grain,
defined by
$$ t_{sp} = a \left| {da\over dt}\right|^{-1}\eqno(15)$$
which can be compared to the
flow time to the core radius, defined by
$$ t_{flow}\left(\xi\right) = r_c~\int^\xi_1 {d{\xi '}\over \left| u\left(
\xi '\right)\right|}.
\eqno(16)$$
These timescales are plotted in Figure 3 for the three galaxy models of
Table 1.  The sputtering time is computed assuming a large grain with a
radius of $0.3\mu {\rm m}$.  Smaller grains have correspondingly shorter
sputtering times.  Figure 3 indicates that sputtering generally occurs
faster than the time for the grains to flow to the core in larger
galaxies, but in smaller galaxies
grains may be carried inward by the cooling flow for a significant
distance before sputtering away.
This suggests that the grain population
at a given radius in the smaller galaxies will have an
admixture of grains that originated further out in the galaxy.  In contrast,
the largest galaxy produces and sputters grains on the spot,
except at the innermost radii where large grains may flow somewhat.

In the case of the largest galaxy (model a),
the grain size distribution in the
ISM can be determined approximately by balancing the local rate
of grain input from stars and the rate of grain destruction by sputtering.
That is,
$$ {\partial\over \partial a}\left( N {da\over dt}\right) = S(a)\eqno(17)$$
where $N$ is the number density of grains per unit grain radius, $S$
represents the stellar
source of grains, and $\dot a = da/dt$ is given by eq. (14).  If, for
example, we assume that the grain size distribution within
recently introduced stellar ejecta
varies as a power law, i.e.
$S\left( a\right) = A a^{-g}$, then eq. (17) is readily solved to give
$$ N(a) = N(a_{max}) + {A\over \bar h\rho}\left(1\over 1 - g\right)\left(
a_{max}^{1-g} - a^{1-g}\right)\eqno(18)$$
where $a_{max}$ is the adopted size of the largest grain.  The constant
$\bar h$ is related to $h$ of eq. (14) by
$$\bar h = {h\over  m_p}\left[ \left({T_d\over T}\right)^\omega
+1\right]^{-1}\eqno(19)$$
where $T$ is the (uniform) temperature of the hot ISM.
Since we assume that there are no sources of grains bigger than
$a_{max}$, $N(a_{max}) = 0$ in eq. (18).  Thus, for
small grains, the size distribution in the ISM
goes approximately as a power
law in the grain size, $N(a) \propto a^{1-g}$,
but with an exponent that differs by unity
from that of the source function.
As $a$ approaches $a_{max}$, the size distribution steepens
and falls towards zero.  Note that because the sputtering rate depends
only weakly on $T$ for the temperatures under consideration, both
$\dot a$ and the grain size distribution are accurately determined
despite uncertainties in our evaluation of the gas temperature (Figure 1).

Eq. (18) demonstrates that sputtering significantly modifies the grain
size distribution from that in the stellar ejecta;
smaller grains are relatively less numerous in the intergalactic medium
compared to that of the stellar ejecta.  The general case where
inward advection of the dust becomes important for determining the grain
size distribution will be treated in \S6, however, the
approximate distribution of eq. (18)
applies for a large region of galaxy model a and the
outer regions of galaxy models b and c. The predicted IR spectrum
and bolometric IR luminosity
are greatly modified due to the reduction in the
dust-to-gas ratio and the altered
grain size distribution as a result of sputtering.
The relative underpopulation
of smaller grains implies decreased emission at smaller IR wavelengths.
\vskip .2in
\centerline{5. SOURCES FOR DUST HEATING}
\vskip .2in
\centerline{5.1 Heating by Ambient Starlight}

Absorption of ambient optical radiation by grains is an
important means of grain heating.
In order to estimate this heating we must
compute the mean intensity of starlight at any radius in the
model galaxies.  The mean intensity at radius $r$ is given by
$$J_\nu\left( r\right) = {1\over 4\pi} \int_\Omega I_\nu\left(r,\Omega \right)
{}~d\Omega = {1\over 2}\int_{-1}^{1}I_\nu ~d\mu_\theta\eqno(20)$$
where $\mu_\theta=cos~\theta$, $\Omega$ is the solid angle,
and $I_\nu$ is the specific intensity of starlight ($\rm ergs~s^{-1}~
cm^{-2}~ster^{-1}~Hz^{-1}$).
To compute $I_\nu$, we assume that the
stellar emissivity $j_\nu$ ($\rm ergs~s^{-1}~cm^{-3}~Hz^{-1}$)
is proportional to the stellar density:
$$ j_\nu\left( \xi\right) = j_o~\left( 1 + \xi^2\right)^{-{3\over 2}}
\eqno(21)$$
where $j_o$ is an as yet unspecified constant which carries all the
frequency dependence of $j_\nu$.  We further assume that dust
distributed throughout the galactic volume is optically
thin to stellar light.  (This latter assumption can be checked for
consistency once the dust content of the interstellar gas is computed in
\S 6.) The specific intensity $I_\nu\left( \xi,\mu_\theta\right)$
in a direction specified by the azimuthal angle $\mu_\theta$
is computed as an integral along a ray starting at the current
location and extending to the outer boundary of the galaxy,

$$I_\nu\left( \xi,\mu_\theta\right) = r_c~\int_0^{l_{\rm max}} j_\nu (\tilde
\xi)~dl.\eqno(22)$$
In this expression, $l$ is a dimensionless length along the line of sight,
$${\tilde\xi}^2 = \xi^2 + l^2 + 2\xi~l~cos\theta,\eqno(23)$$
and
$$l_{\rm max} = -\xi\mu_\theta
+ \sqrt{\xi^2\mu_\theta^2 + {\xi_t}^2 -\xi^2}.\eqno(24)$$

Eq. (22) can be evaluated to yield
$$ \eqalign{ I_\nu\left( \xi,\mu_\theta\right) = j_o~r_c~&\Biggl[{l_{\rm max}
+ \xi\mu_\theta\over \left( 1+\xi^2 - \xi^2\mu_\theta^2\right)
\sqrt{1 + \xi^2 +2\xi\mu_\theta l_{\rm max} + l_{\rm max}^2}}\cr
&-~{\xi\mu_\theta\over \left( 1 +\xi^2 -\xi^2\mu_\theta^2\right)
\sqrt{1 + \xi^2}}\Biggr].\cr }\eqno(25)$$
The mean intensity of starlight is then given by eq. (20),
$$ J_\nu\left(\xi\right) = {j_o r_c\over 2\xi\sqrt{1 + \xi^2}
\sqrt{1 + \xi_t^2}} \Biggl[ \bar
J\left(0,-B\right) - \bar J\left(\xi,-B\right) +
\bar J\left(\xi,B\right) - \bar J\left(0,B\right)\Biggr]\eqno(26)$$
where the function $\bar J$ is defined as
$$ \eqalign{\bar J \left(x,y\right) = \sqrt{A^2+x^2}& - y~{\rm ln}\left(
2\sqrt{A^2+x^2} + 2x\right)\cr
&- \sqrt{A^2 + y^2}~{\rm ln}~\left|{
2\sqrt{A^2 + y^2}\sqrt{A^2 + x^2} -2xy + 2A^2\over x +
y}\right|.\cr}\eqno(27)$$
In these last two equations, $ B^2=1+\xi^2$, and $ A^2=\xi_t^2-\xi^2$.
At the center of the galaxy, the mean intensity is given by
$$ J_\nu\left(0\right) = {j_o r_c \xi_t\over \sqrt{1 +
\xi_t^2}}\eqno(28)$$
and at the outer edge of the galaxy its value is
$$ J_\nu\left(\xi_t\right) = {j_o r_c \over 2\xi_t\sqrt{1 + \xi_t^2}}~
{\rm ln}\left(1 + \xi_t^2\right).\eqno(29)$$

The local heating ($\rm ergs ~s^{-1}$)
of a spherical dust grain of radius $a$ by ambient starlight is
$$ H_* = \int \left({4\pi J_\nu\over c}\right) cQ_\nu \pi
a^2 d\nu.\eqno(30)$$
Here $c$ is the
speed of light, $Q_\nu$ is the dust absorption efficiency (kindly
provided by Dwek 1993), and the integration is over optical and
ultraviolet frequencies.

To evaluate $H_*$, we must determine
the coefficient $j_o$.  Assuming that
the stellar mass to light ratio is spatially constant, the ratio
of the stellar mass density to the stellar emissivity in a given
frequency interval is also constant by assumption (21).
Therefore the ratio of
$\rho_{*o}$ in eq. (1) to $j_o$ in eq. (21), when integrated over the B--band,
is fixed by the stellar mass to B--band light ratio of eq. (A2),
$$ {j_{oB}\over\rho_{*o}} = \left(M_{*t}\over L_B\right)^{-1}.\eqno(31)$$
The determination of $j_{oB}$ allows the total heating due to starlight
in the B--band to be computed.  However, we require the heating due to all
of the starlight, not just the light in the B--band.
This is obtained by multiplying the heating due to
starlight in the B--band with a bolometric correction factor given by
$$ BC={\int_{\rm uv,~ optical} F\left(\lambda\right)Q_\lambda
d\lambda\over \int_{\rm B~ band}F\left(\lambda\right)Q_\lambda
d\lambda}\eqno(32)$$
where $Q_\lambda(a)$ is again the dust absorption efficiency
at wavelength $\lambda$.
The upper integral extends over all optical and ultraviolet wavelengths, the
lower integral extends only over the B--band, and $F\left(\lambda\right)$ is
the spectrum of the stellar light in ellipticals. We assume that
$F\left(\lambda\right)$ does not depend on the radius $\xi$ and we use the
analytic fit
$$ {\rm log}~ F\left(\lambda\right) = \cases{
{\rm log} \left(\lambda\right) - 3.215 & ${\rm if}~ \lambda < 1000{\rm\AA}$\cr
-1.509~{\rm log}\left(\lambda\right) + 4.314 & ${\rm if}~
1000{\rm \AA}\leq\lambda
< 1800{\rm \AA}$\cr
-0.6 & ${\rm if}~ 1800{\rm \AA}\leq\lambda < 2123{\rm \AA}$\cr
4.078~{\rm log}\left(\lambda\right) - 14.17 & ${\rm if}~
2123{\rm \AA}\leq\lambda < 4842{\rm \AA}$\cr
0.86 & ${\rm if}~ 4842{\rm \AA}\leq\lambda < 8913{\rm \AA}$\cr
-0.8812~{\rm log}\left(\lambda\right) + 4.341 & ${\rm
if}~8913{\rm \AA}\leq\lambda < 16255{\rm \AA}$\cr
-2.837~{\rm log}\left(\lambda\right) + 12.58 & ${\rm
if}~\lambda > 16255{\rm \AA}$\cr
},\eqno(33)$$
where $\lambda$ is given in $\rm \AA$ and $F$ is in arbitrary units;
the absolute normalization of
$F\left(\lambda\right)$ is given by eq. (31).
Equation (33) gives a good fit ($\lta 15\%$ in the optical part) to the
spectrum of stellar light in elliptical galaxies as determined by
Bruzual (1985), Renzini \& Buzzoni (1986), and Ferguson \& Davidsen
(1993).
\vskip .2in
\centerline{5.2 Heating by Hot Electrons}

While dust grains are sputtered by collisions with thermal protons and
helium nuclei, collisional heating of grains is dominated by more frequent
collisions with equally energetic but more rapidly moving electrons in
the hot interstellar gas.
At plasma temperatures $T \gta 10^6$K the electric charge of the
grains has a negligible influence on these collision rates
(Draine \& Salpeter 1979).
For larger grains the energy transferred from
the colliding electron to the dust grain is small relative to the
total thermal energy of the grain.  Heating therefore occurs in a smooth
fashion, requiring many electron impacts.
For small grains, however, a single electron collision can introduce
enough energy to significantly raise the temperature of the grain.
For small grains, electron collisions are sufficiently infrequent
that substantial cooling
by radiative losses may occur between successive impacts.
Although small grains are
stochastically heated (Dwek 1986), the time-averaged
heating rate due to hot electrons is given by eq. (3)
of Dwek (1986):
$$H_e = \pi a^2 \left( {\rho \over m_p }\right)\int_0^\infty E
f\left(E\right)v\left(E\right)\zeta\left(E\right) dE\eqno(34)$$
where $E$ and  $v\left(E\right)$ are the energy and velocity of the
impinging electron, and $\rho$ is the gas density.  The
energy distribution of the electrons is Maxwellian,
$f\left(E\right) =
2\pi^{-1/2}\left(kT\right)^{-3/2}$ $E^{1/2}{\rm
exp}\left( -E/kT\right)$, where $T$ is the gas temperature and $k$ is
Boltzmann's constant.  The function $\zeta(E)$ gives the fraction of the
energy of an incident electron that is absorbed by the grain
(Dwek \& Werner 1981),
$$\zeta\left(E\right) = \cases{
1 &${\rm if}~ E<E_*$\cr
1-\left[ 1 - \left(E_*/E\right)^{3/2}\right]^{2/3} &${\rm if}~ E\geq E_*$\cr
},\eqno(35)$$
where $E_*({\rm ergs})=3.7\times 10^{-8}\left(a/\mu {\rm m}\right)^{2/3}$.

\vskip .2in
\centerline{5.3 Heating by X--Rays}

Some fraction of the thermal X--rays emitted by the hot interstellar gas
in ellipticals is absorbed by the dust. Since the
energy contained in each X--ray photon is comparable to the
thermal energy of a grain in equilibrium with the stellar radiation
field, X--rays also cause stochastic heating of the grain.
Stochastic heating can also occur for the smaller grains when an
ultraviolet stellar photon is absorbed (see e.g., Draine \& Anderson
1985).  Although stochastic temperature excursions
are important in considering the predicted emission spectrum from the
grains, we restrict our attention here to the average heating rate due
to X--ray absorption.

The X--ray emissivity of the hot gas is given by
$$ j_X\left(\xi, \nu\right)={\left[\rho\left(\xi\right)\over  m_p
\right]}^2 \Lambda _X\left(T, \nu\right)\eqno(36)$$
where $\Lambda _X$ is the atomic emissivity of the hot gas
($\rm ergs~cm^3~s^{-1}~Hz^{-1}$) and $T$ is the gas temperature.
Since $\rho$ is given by eq. (4), $j_X$ can be represented by
eq. (21) with the identification $j_o=(\rho_o /m_p)^2
\Lambda _X\left(T,\nu\right)$.
The heating rate can then be computed using eq. (30) integrating
over X--ray frequencies.  We assume $Q_\nu=1$ so that
we actually compute an upper bound to the X--ray heating.  The
integral in eq. (30) reduces to an integral over only $j_o$
since the entire frequency dependence of $J_\nu$ is contained in
$j_o \propto \Lambda _X\left(T, \nu\right)$.
\vskip .2in
\centerline{5.4 Comparison of Heating Rates}

The heating rates per grain for each of the three processes discussed above
are plotted as a function of grain size in Figure 4
for the specific case of
graphite grains at a radius of $0.4~\xi_t$ in galaxy model a.
Heating due to starlight is orders of magnitude greater than that
due to either electronic collisions or absorption of thermal
X--rays.  The dominance of heating by starlight
absorption holds at all radii for all galaxy models
in Table 1 and for both graphite and silicate grains.
Heating rate variations with grain size for other radii
and galactic models are qualitatively
similar to those shown in Figure 4 and are not presented.

These results are consistent with previous findings based on less quantitative
treatments (de Jong 1986, Jura 1986) and can be expected from
elementary considerations.  First, for even the brightest
ellipticals, the optical luminosity (B-band) is stronger than X--ray luminosity
by a factor of $\sim 100$ (see eq. [6])
and this factor carries over to grain heating.
Further, unless the energy lost from the hot interstellar gas through
electronic heating of grains is vastly larger than that lost through
plasma radiation, grain heating by electronic collisions must also be small
(the
various channels of energy loss for the hot gas are considered in \S7).

Similar energetic considerations for cD galaxies or for elliptical galaxies
at the centers of clusters of galaxies would be very interesting since it has
been suggested that the principle grain heating mechanism is electronic
collisions rather than optical heating (Bregman, McNamara, \& O'Connell
1990; de Jong et al. 1990).
Based on their far
larger ratios of $L_x$ to $L_B$ (e.g. Edge \& Stewart 1991),
cD galaxies have a much greater amount of
hot gas per star then do isolated elliptical galaxies;
they also have much larger ratios of $L_{IR}$ to $L_x$ (Edge \& Stewart 1991).
The implication is that higher ratios of gas to stellar
density in cD galaxies
favor plasma cooling by electron-grain interactions.
This will be considered in a subsequent paper.
\vskip .2in
\centerline{6. EVOLUTION OF THE GRAIN SIZE DISTRIBUTION}

The equation for evolution of grains in a hot flowing gas with sources
is
$${\partial N\left(a,r,t\right)\over \partial t} + {\partial\over \partial a}
\left[ N\left(a,r,t\right) {da\over dt}\right]
+ \vec\nabla\cdot\left[N\left(a,r,t\right)\vec u\right] =
S\left(a,r,t\right). \eqno(37)$$
In this equation, $N(a,r,t)$ is the grain size distribution at galactic
radius $r$ and is given in units of $\rm cm^{-3}\mu m^{-1}$,
$\dot a = da/dt$ is the sputtering rate given by eq. (14) and $S$
is the source function for dust.  Assuming spherical symmetry and
steady state conditions, this reduces to
$$ -\bar h\rho{\partial N\over \partial a} + u{\partial N\over \partial r} =
S\left(a,r\right) - N\left(2{u\over r} + {du\over dr}\right)\eqno(38)$$
where $u$ is the radial velocity and $\bar h$ is given by eq. (19).
We recognize that the source of grains $S \propto \alpha_*(t)$ is a
strong function of time, but it will be apparent from our results below
that the steady state approximation is sufficient to illustrate the
effects of grain evolution.  The quantity $S(a,r)$ in eq. (38) is therefore
the source function for dust at the present time, $t_{now}$.

The differential equation (38) is quasi--linear and can
be solved by the standard method of characteristics.  The velocity of the
gas $u(r)$ has already been determined by the cooling flow calculations of \S3,
but the source function $S$ and appropriate boundary conditions
need to be specified.
Since we only consider dust ejected from evolving stars, $S$ is simply
given by the rate of dust input from these stars.
Unfortunately, there are presently few
observational constraints on the grain content of red giant winds since
dust absorption features are difficult to observe in the spectra of these
cool stars.  What constraints exist suggest that there are relatively few
small grains ($a<0.08\mu {\rm m}$, Seab \& Snow 1989).  However,
in view of the absence of information on the
grain size distribution we
assume the grain size distribution is similar to that in our Galaxy:
a power law with an exponent
given by the MRN (Mathis, Rumpl, \& Nordsieck 1977) value of $g = 3.5$,
$$S(a,r) = \alpha_n \rho_*\left({A\over \mu m_p}\right)a^{-g},\eqno(39)$$
where $A$ is an as yet
undetermined normalizing coefficient.  We  adopt a minimum grain
size $a_{min} =0.001\mu {\rm m}$ and a maximum grain size of $a_{max} =
0.3\mu {\rm m}$.

In the standard MRN model, the value of $A$ is determined by
extinction measurements of the Galactic interstellar medium.  Since these
data do not exist for our case, we determine $A$ based on the local
metallicity of
stars in elliptical galaxies.  Consider the grain size distribution of an
evolving star:
$$ N_{ej}(a)=\rho_{ej}A_{ej}a^{-g}\eqno(40)$$
where $\rho_{ej}$ is the gas density in the ejecta.  The total mass
density of grains in the ejecta is then
$$ \rho_{dust} = \int^{a_{max}}_{a_{min}} \left({4\over 3}\pi a^3
\rho_{grain}\right)~ \rho_{ej}A_{ej}a^{-g} da\eqno(41)$$
where $\rho_{grain}$ is the mass density of an individual grain.  We take
$\rho_{grain}=2.2~ \rm g~cm^{-3}$ for graphite and $\rho_{grain}=3.0~
\rm g~cm^{-3}$ for silicate.
The density of dust is related to the
gas density by the dust to gas ratio $y_g$:
$$ \rho_{dust} = y_g\rho_{ej}.\eqno(42)$$
Then, from eq. (41):
$$ A_{ej}={y_g\over {4\over 3}\pi\rho_{grain}}\left(\int^{a_{max}}
_{a_{min}} a^{3-g}~da\right)^{-1}.\eqno(43)$$
The solar abundance by mass of both silicon and carbon is about 0.005.
We therefore set
$$ y_g=0.005~\delta~Z\left(r,L_B\right),\eqno(44)$$
where $\delta$ is the fraction of either Si or C in the ejecta contained
in grains, and $Z$ is the metallicity of the stars relative to
solar.  Based on optical line width measurements
(Davies, Sadler, \& Peletier 1993; Schombert et al. 1993; Gonzales et al.
1994),
the metallicity is known to vary
with radius and B--band luminosity of the galaxy approximately as
$$ Z\left(r,L_B\right) = {\left(L_B/10^{10}L_\odot\right)^{0.301}\over
1 + \left(r/r_c\right)^{0.3}}\eqno(45)$$
where $r_c$ is the galaxy core radius.  We therefore find that
$${A\over \mu m_p}=A_{ej}=\left({y_g\over {4\over 3}\pi\rho_{grain}}\right)
\left({4-g\over a_{max}^{4-g} - a_{min}^{4-g}}\right)\eqno(46)$$
where $y_g$ is given by eq. (44).
In deriving $S$ from (39) we have assumed that few grains are sputtered
before the stellar ejecta merges into the hot ISM.
We provide a brief argument in Appendix C to support this
assumption.

It is useful to compare our values for the normalization $A$ with
standard values associated with the MRN distribution.  For
carbon grains in a typical
galaxy of $10^{11} L_\odot$, we find $A=0.053$ at the
core radius $r_c$.  This assumes that all the carbon in the ejecta is
incorporated into
grains ($\delta_{C} = 1$).
This is close to the standard MRN value of $A=0.069$
(Draine \& Anderson 1985) for graphite grains in the Galaxy.
For silicate, again assuming $\delta_{Si} = 1$,
we get $A=0.039$ which is similar to the MRN value
$A=0.078$.
Although our adoption of the source function
in eqs. (39 - 46) is approximate,
a detailed comparison of infrared observations with computed
IRAS spectra will lead to an improved understanding of the grain
abundance and size
distribution in the ejecta of red giants and planetary nebulae.

Regarding an appropriate boundary condition at large galactic
radius for solving eq. (38),
we note that a favorable aspect of our method
is that the grain size distribution in the galaxy actually
does not depend on
the choice of the outer boundary condition although something
has to be specified for the numerical calculation.  This is seen by
referring to the timescales of Figure 3.  In all cases, towards the outer
boundary, the flow time far exceeds the sputtering time of the largest
assumed grain.
Indeed, the flow time tends to infinity as the flow velocity decreases
toward the outer boundary.
The grains
in the outer regions therefore sputter on the spot and the grain size
distribution at small or intermediate galactic
radii is insensitive to the distribution
at larger radii.  We may specify any grain size distribution
at large radius to start the integration of the grain flow equation
(eq. [38]).  For consistency, we assume that the grain content
at large radius is determined by the on the spot sputtering assumption of
eq. (18) with
the source function of eq. (39), i.e.
$$ N\left( a,r\right) = {\alpha_n\rho_*\over \left(g-1\right) \bar h\rho}
\left({A\over \mu m_p}\right)\left(a^{1-g} - {a_{max}}^{1-g}\right).
\eqno(47)$$
Finally, we must specify that the number density of grains vanish
for the biggest grains
in the ISM $a_{max}$, $N\left( a_{max} ,r \right)=0$ at all $r$.

Figure 5a shows grain size distributions for graphite grains determined
from eq. (38) for model
galaxy a.  The grain size distributions are given at the core radius $r_c$
(topmost curve), and at successively larger radii (lower curves).  The
first implication of this figure is that the grain size distribution
is very close to that of {\it in situ} sputtering (eq.
[18]).  For small grains the distribution is essentially a power law
that is shallower than that of the stellar ejecta by unity.  At large
grain sizes, the density steepens and drops to zero at the largest assumed
grain size.  We anticipated this result wherever the sputtering times are
shorter than the flow times (Figure 3).  What is
somewhat surprising, however, is that the assumption of {\it in situ}
sputtering
can be used to accurately determine the size distribution even
where the sputtering time for the largest grains is comparable to
or exceeds the flow times to the core.
This can be understood since the flow of grains from upstream
into a given volume $4 \pi r^2 dr$ near the galactic core
is significantly smaller than the amount
of grains produced by stars within $4 \pi r^2 dr$ -- this is a result of
the steep stellar density profile $\rho_*(r)$.

The grain size distributions for graphite in the smallest galaxy
(model c) are shown in
Figure 5b.  In contrast to galaxy model a, the flow time to the core
radius is comparable to the sputtering time for the largest grain over
most of the galaxy.  Again, the grain size distributions are well
determined by assuming {\it in situ} sputtering.  Similar
results hold for galaxy model b which are not illustrated.
Grain size distributions for silicate dust
have identical shapes to those of graphite, although the normalizations are
slightly different, so these results are not shown.  A favorable
consequence of the above results is that the grain size distribution
in the ISM depends only on the ratio of the stellar density to the gas
density (and on the rate of dust injection into the ISM).  These quantities
are reasonably well known.  The grain size distribution does not depend
on the more poorly known gas velocities in the cooling flow.

In order to gain further insight, we have varied $a_{min}$ and $a_{max}$
while keeping the slope of the grain distribution $g$ fixed.
For example,
if instead of $a_{max}=0.3\mu {\rm m}$ we set $a_{max}=1.0\mu {\rm m}$
in the stellar
ejecta, the grain size distribution is modified in the manner shown in Figure
5c.  Over the common range of grain sizes, the case with $a_{max}=1.0\mu
{\rm m}$
gives about half the grain density relative to the case with $a_{max}=0.3
\mu {\rm m}$.  Since the total mass of grains ejected
by stars is fixed, extending the range over which grain masses
are distributed
requires a decrease in the density per unit grain size.  That is,
by increasing $a_{max}$ we decrease the normalization $A$
by a factor of $\sim 0.53$.

One final case we consider is motivated by the observational evidence that
there may be fewer small grains in stellar ejecta than in
the Galactic ISM.
For example, consider a power law grain size distribution
$S(a) \propto a^{-g}$ with
$a_{min} = 0.08\mu {\rm m} < a <
a_{max}=1.0 \mu {\rm m}$.
In the cooling flow ISM, however,
we continue to follow the evolution of grains smaller
than $a_{min}$ since these smaller grains
are produced by sputtering.  The resulting grain size distribution is
shown in Figure 5c as the dashed line.  Since the range of grain sizes
in the source function $S(a)$
is restricted at small grain radius $a$, the density per unit grain
size is higher than in the case where $a_{min} = 0.001\mu {\rm m}$ and
comparable to that of the first case considered (solid line of Figure 5c).
For $a < a_{min}$ the grain size distribution in the ISM is
essentially constant.  Because there are no stellar sources of dust
in this range, the only sources are the sputtered remnants of larger
grains.  Since the sputtering rate is independent of the grain size
(eq. [14]), all grains get smaller at the same rate.  Similar results
are obtained for galaxy models b and c but are not illustrated.
\vskip .2in
\centerline{7. ENERGY LOSS FROM THE INTERSTELLAR GAS}

Given the grain size distributions of \S6, we can compute the total
rate of energy loss from the hot ISM by electron-grain collisions.
This can be done at any given radius by simply convolving the heating
rate per grain (eq. [34]) with the grain size distribution.  This rate of
energy loss from the plasma by electron-grain collisions
can be compared with the
local rate of energy loss due to thermal emission,
$$ {dE_{X}\over dt} = {\left[\rho\left(\xi\right)\over m_p
\right]}^2 \Lambda_{tot} \left(T\right),\eqno(48)$$
where $\Lambda_{tot} \left(T\right)$ is the bolometric atomic
X--ray emissivity.

Cooling rates for the hot gas in the largest galaxy (model a) are shown
as a function of galactic radius in Figure 6.  In all cases, the
energy loss rate from the hot plasma due to grain heating is
significantly smaller than that of the thermal emission.  This result
is very different
from that of Dwek (1987) who found that dust heating accounted
for as much two orders of magnitude {\it more}  energy loss from hot,
dusty plasmas than thermal emission.
Dwek then applied these results to supernova remnants.
The difference between our results and those of Dwek's is our
inclusion of sputtering losses.
To estimate the magnitude of the reduction of the dust-to-gas
ratio due to sputtering, assume that all grains have the
same radius $a$ and mass $m_g$
and that they are created and destroyed at the same rate.
In this case the number density of grains $N$ is related
to the dust-to-gas ratio in the stellar ejecta $y_g$ by
$N m_g/t_{sp} = \alpha_n \rho_* y_g$, where
$t_{sp} \approx a m_p / h \rho$ is the local sputtering time in
plasma of density $\rho$ (eqs. [14] and [15]).
When sputtering is included the dust-to-gas ratio
is $y_{gs} = m_g N / \rho \approx y_g t_{sp} \alpha_n \rho_* / \rho$.
It follows that sputtering reduces the dust-to-gas ratio
by a factor $y_{gs} / y_g \approx t_{sp} \alpha_n \rho_* / \rho
\approx (a m_p \alpha_n / h ) \rho_*/\rho^2$.
Consider for example
stellar and gas densities near the core radius of galaxy
model a, $\rho_*\sim  3 \times 10^{-21}$ g cm$^{-3}$ and
$\rho/m_p \sim 0.1$ cm$^{-3}$.
At this location the dust-to-gas ratio is reduced
by $ y_{gs} / y_g \approx 3 \times 10^{-3}$ for grains of size
$a = 0.01$ $\mu m$; this is essentially the factor that accounts
for the difference in our plasma cooling rates and those of Dwek (1987)
who assumed an ``extended" MRN distribution with an
unsputtered dust--to--gas mass ratio.
By allowing the MRN distribution in the stellar ejecta to be
sputtered,
we find that the plasma energy
loss to grain heating is less, not very much
greater, than radiative losses.
We also find that the space
density of the largest grains is preferentially decreased;
these are the grains
most responsible for energy loss from the hot gas (see below).
Self-consistent grain sputtering therefore plays a decisive
role in correctly assessing the significance of dust
emission for the hot ISM in galactic cooling flows.

It is evident from Figure 6
that the heating of graphite grains (short dashed--dotted
line) accounts for a little more than half of the total plasma cooling
rate whereas the less numerous silicate grains (long dashed--dotted
line) contribute somewhat less than half of the total.  We also find
that the various assumptions made concerning the grain size distribution
of the stellar ejecta do not result in large differences in the rate
of heat loss from the hot ISM.  This is easy to understand by
considering the dependence of the plasma cooling rate
on the assumed grain distribution of the stellar
ejecta.  The heating rate for the dust is
$$ {dE_{dust}\over dt} = \int^{a_{max}}_{a_{min}} N\left(a, r\right)
H_e da~\propto \left({1\over a^{4-g}_{max} - a^{4-g}_{min}}\right)
\int^{a_{max}} _{a_{min}} a^{1-g} a^2 da~ \sim ~\rm Constant.\eqno(49)$$
The term in parenthesis is due to the normalization of
the grain size distribution (eq. [46]), the first factor in the second
integrand is roughly the power law of the grain size distribution in the hot
ISM, and the second factor in the second integral is approximately
proportional
to the cross-section of the grain for electron-grain collisions.
The plasma cooling rate due to electron-grain collisions is
therefore insensitive
to the assumed value of the slope $g$ or the
upper and lower cutoffs in the grain size distribution
in the stellar ejecta.
Eq. (49) shows
that the plasma cooling rate depends only on the total mass of grains
ejected by stars into the ISM and the ratio of the stellar mass to
gas mass.  Since these are well constrained by observations,
the rate of plasma cooling by this process is well determined
regardless of specific assumptions made about the grain size
distribution in the stellar ejecta.

According to the results plotted in Figure 6, the ratio of energy
loss from the hot ISM by dust heating to that by X--ray
emission is relatively independent of galactic radius $r$.
The heating of individual
grains by hot ISM electrons is proportional to the local gas
density, $H_e \propto \rho$ (eq. [34]).  But the local grain density
depends on the ratio of stellar to gas density and the stellar metallicity,
$N(a,r) \propto Z \rho_* / \rho$; therefore $N H_e \propto Z \rho_*$.
By comparison the rate that the hot ISM loses energy by thermal radiation
$dE_X /dt \propto \rho^2 \propto \rho_*$ (eqs. [1], [4], and [48]).
Therefore the ratio of thermal losses by grain heating to optically
thin emission varies slowly with galactic radius,
$N H_e / (dE_X/dt) \propto Z(r) \propto r^{-0.3}$, which
accounts for the small rate of divergence between the two
rates plotted in Figure 6.  For the largest galaxy, for example,
$(r_t/r_c)^{0.3}\sim 159^{0.3}= 4.6$ is the ratio of the two
heating rates at the core radius and at the outer radius,
again consistent with Figure 6.

The loss of thermal energy from the hot ISM by grain heating in
the two smaller galaxies b and c is even less important relative
to radiative losses than in model a. This can be understood
because $N H_e / (dE_X/dt) \propto Z \propto L_B~^{0.3}$ (eq. [45]).
The larger ratio of sputtering to flow times in smaller galaxies
(Figure 3) plays a very minor role in this ratio; the
relative contribution of advected dust to the total dust content at
any radius is small because of the steep radial dependence of $\rho_*(r)$.

\vskip .2in
\centerline{8. CONCLUSIONS}

We have investigated the evolution and energetics of dust embedded in the
hot ISM of isolated elliptical galaxies.  For this purpose we
constructed a series of single-parameter galaxy models assuming
King type profiles for the stellar density which are consistent
with the fundamental plane relation, the mass to light ratio, core
radius relation, and a virial condition for isothermal cores.
We consider three representative model ellipticals
with $L_B$ spanning a range from $\sim 10^{10} L_\odot$
to $\sim 10^{11} L_\odot$ which are expected to contain galactic
cooling flows.

The galaxy models were used to construct approximate isothermal --
but fully time-dependent -- cooling flow models for the hot ISM.
The radial gas density variation is known if the X--ray surface
brightness distributions of all ellipticals resemble those resolved
by {\it Einstein}.
With this assumption we can solve for the velocity field using the
continuity equation alone.  These ``continuity'' models compare well with
detailed hydrodynamical cooling flow calculations.  The isothermal
temperature of the gas is taken to be the same as the mean
stellar temperature of the galaxy.  This latter temperature is determined
by solving the equation of stellar hydrodynamics in the presence
of a massive dark halo.  These temperatures agree reasonably well with
observed temperatures for galaxies of comparable luminosity (see Figure
1).  A more detailed agreement of predicted and observed temperatures
is not required for the purposes of this paper because the results depend
only weakly on the assumed temperature of the ISM.

Having specified the plasma environment within the model galaxies, the
radiation
environment from the near IR through the UV is taken from observations.
With this information we can compute the heating rate of
interstellar grains due to absorption of
stellar light, impacts with energetic electrons from the hot ISM,
and absorption of thermal X--rays from the hot ISM.
In spite of the efficiency of sputtering in grain destruction, the
heating of grains in all model galaxies is dominated by
absorption of ambient starlight by over an order of magnitude.  This result
applies for both graphite and silicate grains.
Although this result was anticipated by de Jong (1986) and Jura (1986),
our detailed calculation of the heating rates is relevant to
the well constrained radiation--plasma environment of elliptical galaxies.
In future work, we will consider the same issue for cD galaxies and
giant ellipticals at the centers of clusters of galaxies where both
the plasma environment and IR emission characteristics appear to be
distinct from isolated ellipticals (see \S5.4).

We have computed the grain size distribution in three
representative galactic cooling flows
including the effects of grain sputtering,
advection by the cooling flow, and local stellar sources of dust.
In this paper we have not considered external sources
of grains -- for example by recent mergers of gas-rich galaxies
with ellipticals -- or dust that may form within gas
deposited in the centers of the cooling flows.
In this sense our estimate of the amount of dust present
in ellipticals is a lower limit.
We assumed that the grain size distribution introduced into the
cooling flow at every radius from evolving stars is given by
a power law having the MRN index of $g = 3.5$, normalized according to the
observed local metallicity of the parent stars.
No significant loss of grains is expected
during the brief transition from red giant winds or planetary
nebulae into the ISM.
The ratio of grain sputtering time to radial flow time varies within each
galactic cooling flow and among galaxies of different total mass.
When the sputtering time is very short we find that the
grain size distribution for small grains will be a
shallower power law with index $1-g$,
but we find that this same solution is still an excellent
approximation even when the sputtering time is so long
that the grains can move inward during their lifetimes
across an appreciable part of the galaxy.
This last more surprising result can be understood from the
very steep radial gradient of the stellar sources of grains;
grains advected from larger radii in the galaxy are
far less numerous compared to those
produced locally at every radius in the cooling flow.
This latter property implies that the grain size distribution does not
depend on the fluid velocities in the cooling flow.  Rather, the distribution
depends only on the better known ratio of the stellar density to the
gas density.

The heating of grains by absorption of hot thermal electrons is a
possible channel of energy loss for the hot gas in cooling flows.
Dwek (1987) suggested that this mode of energy loss may far exceed that
due to emission of thermal X--rays in hot plasma environments similar
to that of the ISM of ellipticals.  However, we find that this channel
for energy loss is in fact much smaller than that due to thermal emission
in galactic cooling flows.  While Dwek's result was based on assuming
that the grains (with an MNR distribution)
are unsputtered while they reside in the hot plasma,
in our model galaxies the dust-to-gas ratio and the distribution of grain sizes
are considerably modified by sputtering.
The total plasma cooling by a combination of graphite and
silicate grains is a factor $\gta 5$ below that of optically
thin radiative losses in all galaxies we consider.
Our results for the magnitude of plasma
cooling by thermal interactions with grains
are very insensitive to the assumed upper and lower grain
size cutoffs as well as the slope $g$ of the grain distribution in the
stellar ejecta.

In this paper we have surveyed the relevant physics of distributed
grains in elliptical galaxies.
A second paper, currently in preparation, will describe in detail
the temperatures of grains at every radius in elliptical galaxy cooling flows
and the far infrared emission into each of the IRAS bands.
We will determine the radial distribution of
IR emission for comparison with future,
high resolution observations.
We are also interested in estimating the IRAS luminosities from
more centrally concentrated dusty patches and lanes in ellipticals
for comparison with the more distributed dust.

We are grateful to Eli Dwek for providing us with grain absorption
coefficients, to Scott Trager and Sandra Faber for insights
on observations of ellipticals, and to John Raymond for use
of his code. JCT was supported by an NRC associateship
and WGM thankfully acknowledges support from CalSpace and a Faculty
Research Grant from UCSC.
\vfill\par\eject
\centerline{APPENDIX A}
\centerline{SELF-CONSISTENT GALAXY MODELS}
\vskip .2in
The fundamental plane for elliptical galaxies
$$ L_B = 3.6 \times 10^3~ R_e \sigma_{*o}^{1.3} h^{-1}\eqno(A1)$$
relates the B-band luminosity $L_B$ (in units of $L_{\odot B}$), the
effective radius $R_e$ (in parsecs) and the central stellar velocity
dispersion $\sigma_{*o}$ (in km s$^{-1}$); $\rm h = H/100~ Mpc~km^{-1}~
s^{-1}$ is the Hubble parameter (Dressler et al. 1987; Djorgovski
\& Davis 1987; Ciotti et al. 1991).  The stellar luminosity $L_B$ can
be directly related to the total stellar mass $M_{*t}$ (in units of
$M_{\odot}$) by the stellar mass to light ratio:
$$ M_{*t}/L_B = 2.98 \times 10^{-3}~ L_B^{0.35} h^{1.7} \eqno(A2)$$
based on the recent study of van der Marel (1991).

The effective radius $R_e$ that contains half of the stellar mass or
light in projection can be found from eq. (3) in terms of $r_c$ and
$r_t$.  The resulting implicit equation for the effective radius
$\lambda_e = R_e/r_c$ is approximated very well by
$$\lambda_e = 0.857~ \xi_t^{1/2} \eqno(A3)$$
where $\xi_t = r_t/r_c$.  In addition, the core radius $r_c$, in parsecs,
increases with galactic luminosity according to
$$r_c = 6.84 \times 10^{-11}~ L_B^{1.20} h^{1.4} \eqno(A4)$$
(Lauer 1989).  This relation is consistent with more recent galactic
distances based on the $D_n - \sigma_*$ method (Donnelly, Faber, \&
O'Connell 1990).

Finally, the core radius, central density, and central velocity dispersion
are related via the standard core relation for isothermal spheres (Binney
\& Tremaine 1987),
$$r_c = \sqrt{9\sigma_{*o}^2\over 4\pi G \rho_{*o}}~.\eqno(A5)$$
By elimination among eqs. (A1)--(A5), all quantities of interest can be
expressed in terms of a single parameter, most conveniently chosen to be
the total radius $\xi_t$.  For a given $\xi_t$, the optical luminosity (in
$L_{\odot B}$) is given by
$$ L_B = 8.92 \times 10^{11}~ \xi_t^{-1.682} \mu_{mass}(\xi_t)^{2.186}
h^{-2} \eqno(A6)$$
where $\mu_{mass}$ is given by eq. (2) of the text.
The central stellar density is then given by
$$ \rho_{*o} = 4.41 \times 10^{10}~ \xi_t^{-0.769}
L_B^{-2.708} h^{-3.416}~~~ \rm{g~cm}^{-3}. \eqno(A7)$$
The corresponding core radius and total stellar mass are found from
eqs. (A4) and (A2), respectively.  Table 1 lists the parameters of
a representative sample of the galaxy models generated by the above
procedure.  We assume $h=0.75$ throughout.

We now estimate the mean stellar temperature for our galaxy models.
Gas ejected from stars is heated to the virial temperature by dissipation.
The gas can be further heated by Type Ia supernovae if the supernova rate
is sufficiently large. The observed supernova rate is usually given in units
of SNU = 1 supernova every 100 years in a galaxy of $L_B = 10^{10}$
$L_{\odot B}$.  For a typical large elliptical galaxy supernova heating
dominates heating by stellar motions if SNU $\gta ~ 0.1$.  However, recent
observations of NGC 1399 and NGC 4472 by Serlemitsos et al. (1993)
and Forman et al. (1993) indicate iron abundances in the cooling
flow gas that are near solar, similar to that of the stellar ejecta.
The implied value of SNU is less than 0.05 when compared with the models
of Loewenstein and Mathews (1991).  If supernova heating is unimportant,
as seems likely, the gas temperature should be close to the mean stellar
temperature $\langle T_* \rangle$.  We determine approximate values for
$\langle T_* \rangle$ by solving the equation of stellar hydrodynamics
for the velocity dispersion (Mathews 1988) with plausible assumptions for
the dark halo mass distribution and stellar velocity ellipsoids.  The
stellar temperature is then determined from the velocity dispersion by
$$\sigma^2_*(\xi) = {3 k T_*(\xi)\over \mu m_p} = {3 \over 2}{G M_{c*}
\over r_c} \tau(\xi), \eqno(A8)$$
where $\mu m_p$ is the mean mass per particle ($\mu$ is taken to be 0.5
and $m_p$ is the proton mass), $M_{c*} = 4 \pi r_c^3 \rho_{*o}/3$
is the galactic core mass,  and $\tau$ is a dimensionless temperature.

We assume that dark halos in ellipticals are pseudo-isothermal
$$\rho_h(\xi) = \rho_{ho} [1 + (\xi/\beta)^2]^{-1}\eqno(A9)$$
with a universal ratio of halo to stellar core radii $\beta = r_{ch}/r_{c}
= 10$ and having the same outer radius $\xi_t$ as the stellar distribution.
The total integrated masses of the dark halos is assumed to be larger than
$M_{*t}$ by a factor $\varpi = 10$.  This implies values of $\rho_{ho}$
listed in Table 1.  For the stellar velocity ellipsoid
we adopt a radial dependence given by
$$ b(\xi) = 1 - { {\bar{v_{tr}^2}} \over
{\bar{v_{r}^2}} } = \left({ \xi \over \xi_t}\right)^{2}\eqno(A10)$$
(Mathews 1988) where $\bar{v_{tr}^2}$ and $\bar{v_{r}^2}$ are the
transverse and radial velocity dispersions respectively.

The stellar temperatures weighted by stellar mass and
listed in Table 1 are computed
from the temperature distribution $T_*\left( \xi\right)$ using
$$\langle T_* \rangle = T_{*o} {\bar{\tau}}
= T_{*o} { \int_{0}^{\xi_t} \tau(\xi) \rho_*(\xi) \xi^2 d \xi
\over \int_{0}^{\xi_t} \rho_*(\xi) \xi^2 d \xi }~;~~~~ T_{*o}\equiv
{\mu m_p\over 2k}~{GM_{c*}\over r_c}.\eqno(A11)$$
We find that the average temperatures computed this way imply the relation
$\langle T_* \rangle \propto L_B^{0.54}$ or equivalently,
$L_B \propto \langle \sigma_*\rangle^{3.7}$.  Since this is close to the
observed Faber--Jackson law, we conclude that the assumed properties for
the massive halo and velocity ellipsoid are reasonable.  In addition,
recently measured temperatures from ASCA
(Serlemitsos, et al. 1993; Mushotzky et al. 1995)
agree well with eq. (A11) and
our relation $\langle T_* \rangle \propto L_B^{0.54}$ (Figure 1).
\vskip .2in
\centerline{APPENDIX B}
\centerline{THE $L_x,L_B$ RELATION}
\vskip .2in
Canizares, Fabbiano \& Trinchieri (1987) showed that the X-ray luminosity
in the {\it Einstein} IPC band ($\Delta E \sim$ 0.5 - 4.5 keV) scales with
optical luminosity as $L_x \propto L_B~^{1.5 - 1.7}$.  Later Donnelly,
Faber, \& O'Connell (1990) found $L_x \propto L_B~^{2.0 \pm 0.1}$ using
a smaller sample but with better galactic distances ($D_n - \sigma_*$
method).  Sarazin (1987) showed that $L_x \propto {\dot{M_{*t}}} \sigma_
*^2$ provided the cooling flow gas is heated by compression in the
gravitational field; from this and the Faber-Jackson relation it follows
that $L_x \propto L_B^{1.5}$.  In the following we present an alternate
derivation of this result.

Suppose that the hot interstellar gas in all bright ellipticals has a
homologously similar distribution.  If $\bar{n}$ is the mean plasma density
then the rate that energy is emitted into the {\it Einstein} band per
unit volume is proportional to $\bar{n}^2 \Lambda_{\Delta E}$
($\Lambda_{\Delta E}$ is the atomic emissivity in the {\it Einstein}
band).  The specific emissivity is $\propto
\bar{n} \Lambda_{\Delta E} /\mu m_p$, where $\mu m_p$ is the mean mass
per particle.  The total observed X-ray luminosity should be $L_x
\propto
\bar{n} \Lambda_{\Delta E} M_g/\mu m_p$ where $M_g$ is the total mass of
hot gas.  If $\langle \alpha_* \rangle$ is the specific mass loss rate from
an evolving population of old stars (Mathews 1989) averaged over a cooling
flow time (several Gyr), then by continuity of gas flow $M_g / t_{cool} \approx
\langle \alpha_* \rangle M_{*t}$.  The cooling time is given by $t_{cool}
\approx 5 kT / 2 \Lambda_{tot} \bar{n}$ where $\Lambda_{tot}$ is the
bolometric cooling function.

Combining these results, we expect
$$ L_x \propto \langle \alpha_* \rangle (\Lambda_{\Delta E}
/\Lambda_{tot}) M_{*t} T \propto \langle \alpha_* \rangle (\Lambda_{\Delta E}
/\Lambda_{tot}) M_{*t} L_B^{2/\eta}, \eqno(B1)$$
where we have used the Faber-Jackson law $L_B \propto \sigma_*^{\eta}$ and
the virial condition that $\sigma_*^2 \propto \langle T_* \rangle \approx T$.

If the stellar population in all ellipticals is dominated by equally old
stellar populations, $\alpha_*(t_n)$ should not vary with $L_B$.  However,
eq. (B1) involves $\langle \alpha_*(t) \rangle$ which is averaged
over the time for gas to flow through the central regions ($\xi ~\lta ~
\lambda_e$) where most of the X-rays are produced.  Typically this time is
very short ($\lta ~ 10^8$ yrs; see Fig. 2) so we conclude that $\langle
\alpha_*(t) \rangle \approx \alpha_*(t_n)$ does not vary appreciably with
$L_B$ as long as all ellipticals have about the same age.
For plasmas with solar abundance, $\Lambda_{\Delta E}/
\Lambda_{tot} \approx 0.6$ and is independent of $T$ provided $T ~\gta
{}~ 3 \times 10^6$ K; $\Lambda_{\Delta E}/\Lambda_{tot}$ should not vary
with $L_B$ even for the least luminous galaxy in Table 1.  However, the
stellar mass to light ratio varies as $M_{*t}/L_B \propto L_B^{\nu}$ where
$\nu \approx 0.2$ (Dressler et al. 1987; Djorgovski \& Davies 1987;
Bender, Burstein, \& Faber 1992; Renzini \& Ciotti 1993) or $\nu \approx
0.35 \pm 0.05$ (van den Marel 1991).  Using this latter value and $\eta
= 4$ we find $L_x \propto L_B^{1.9}$, very close to the observed relation.

Finally we note from eq. (B1) that $L_x(t) \propto \langle \alpha_*(t)
\rangle \propto t^{-1.35}$ which accounts for the slow secular decrease
in X-ray luminosity during the cooling flow evolution of ellipticals
(Loewenstein \& Mathews 1991; Ciotti et al. 1991).
\vskip .2in
\centerline{APPENDIX C}
\centerline{DUST GRAINS IN STELLAR EJECTA ENTERING THE COOLING FLOW}
\vskip .2in
The complex evolution of gas ejected from red giants
as winds or planetary nebulae as it interacts with ambient hot
cooling flows has been discussed by Mathews (1990).
A typical evolutionary sequence is: attainment of hydrostatic
equilibrium with the hot gas, disruption
by Rayleigh-Taylor instabilities into many small clouds
that extend back along the stellar orbit as the orbital
motion inherited from the parent star is dissipated,
and conductive evaporation into the hot ICM
after coming to rest.
Gas ejected from stars interacts only with the hot gas
(not with other stellar ejecta) and can
be kept ionized by galactic UV radiation prior to
conductive evaporation.
However, gas ejected from stars
must be conductively heated to cooling flow
temperatures $T \sim 10^7$K in less than $\sim 10^6$ yrs or
the collective emission line luminosity would exceed that observed
in elliptical galaxies.
Conduction of thermal energy from the hot ISM into the
stellar ejecta can occur by magnetic reconnection as soon as the
debris of the stellar ejecta slows below the local Alfven velocity.

In this paper we assume that dust grains survive this
transient evolution from stellar ejecta into the cooling flow gas.
The most convincing argument for this is to adopt
the opposite point of view, a worst case
scenario in which the grains are
destroyed before encountering the hot ISM.
Suppose, for example, that gas ejected from a star remains intact
(i.e. no instabilities!)
and that the density and temperature in the stellar ejecta are
sufficient to allow sputtering destruction of grains
before the gas density drops to ambient ISM values.
Since sputtering is very slow unless $T \gta 10^6$K, we suppose
that the stellar ejecta, while still rather dense,
is heated by conduction to the
ambient temperature $T_a$ of the local cooling flow.
In this (very unlikely) circumstance the grains could sputter
away before entering the hot ISM.
We now consider the self-consistency of this worst case evolution.

For plasma temperatures $T \gta 10^6$K
the sputtering time $t_{sp}$ for grains of size
$a = 10^{-5}$ cm depends
mostly on the plasma density in the cloud $n_c$,
$t_{sp} = a/{\dot a} \approx 1 \times 10^5 a_{-5} n_c^{-1}$ yrs, where
$a_{-5}$ is the grain size expressed in units of $10^{-5} \rm cm$
($n_c$ is in units of $\rm cm^{-3})$.
A spherical cloud of radius $r_c$
with the mass of an ejected stellar envelope,
$M_{ej} = 0.2$ $M_{\odot}$ and temperature $T_c$ will expand
into the low pressure ISM in approximately the sound crossing
time, $t_{sc} = r_c/c_s \approx 2.5 \times 10^3 T_{c7}~^{-1/2}
n_c~^{-1/3}$ yrs ($T_{c7}$ is the temperature expressed in units
of $10^7\rm K$)
where the cloud is assumed to have
been heated to temperatures comparable with the local ISM.
In order for the grains to sputter away before entering the
local ISM, we require $t_{sc} > t_{sp}$ or
$n_c > 246 a_{-5}~^{3/2} T_{c7}~^{3/4}$ cm$^{-3}$.
Since the internal pressure in such a cloud,
$n_c T_c \sim 10^9$ cm$^{-3}$K,
far exceeds that in typical cooling flows,
$n_a T_a \sim 10^4$ cm$^{-3}$K,
the spherical cloud of stellar ejecta will indeed expand.

However, the thermal energy necessary to heat
the stellar ejecta to temperature $T_c$ must
come from the local ISM.
The radius of a region in the local ISM that contains the
same thermal energy as the cloud is
$r_a = r_c (n_c T_c / n_a T_a)^{1/3}
\approx 3.8 \times 10^{19}
T_{c7}~^{1/3} (n_a T_a)_4~^{-1/3}$ cm, where $(n_a T_a)_4$ is $(n_a
T_a)$ given in units of $10^4\rm cm^{-3}K$.
Finally, thermal heat must be conductively transported
into the stellar ejecta before the cloud can expand under
its own pressure.
Therefore in order to destroy the dust before it enters
the local ISM we require that
heat is conducted into the spherical cloud with a characteristic
velocity
$u_{fl} > r_a / t_{sc} \approx 4.8 \times 10^3 n_c~^{1/3} T_{c7}~^{5/6}
(n_a T_a)_4~^{-1/3}~ {\rm km~s^{-1}}>
3 \times 10^4 a_{-5}~^{1/2} T_{c7}~^{13/12}
(n_a T_a)_4~^{-1/3}$ km s$^{-1}$.
Since this lower limit
exceeds the velocity of thermal electrons in a
gas at $T_a \sim 10^7$K,
$v_e \approx 2 \times 10^4 T_7~^{1/2}$ km s$^{-1}$,
we conclude that this hypothetical, maximally pessimistic
evolutionary model is impossible.
Dust grains do in fact largely survive the transition from
stellar ejecta into the local cooling flow ISM
as we have assumed here.

\vfill\eject
{
\def\skip#1{\noalign{\vskip #1pt}}
\def\lline{\noalign{\hrule}}

$$\vbox{
\halign to 6.3truein{#\hfill\tabskip=10pt plus 20pt minus 20pt&\hfill#\hfill
&\hfill #\hfill&\hfill #\hfill&\hfill #\hfill
&\hfill #\hfill&\hfill #\hfill&\hfill #\hfill\tabskip=0pt\cr
\multispan8\hfill TABLE 1\hfill\cr
\skip{9}
\multispan8\hfill Galaxy Model $\rm Parameters^a$\hfill\cr
\skip{6}
\lline
\skip{2}
\lline
\omit\vrule height 14pt width 0pt \hfill Model\hfill&
$L_B (10^{10}L_\odot)$&$r_c (\rm kpc)$&$r_t (\rm kpc)$
&${\rho_{*o}}^b$&${\rho_o}^b$
&${\rho_{ho}}^b$&
$\langle T_*
\rangle (10^7 \rm K)$\cr
\skip{5}
\lline
\vrule height 14pt width 0pt%
a&10.6&0.774&123.&$3.37$&$2.42\times 10^{-4}$&$1.12
\times 10^{-2}$&0.922\cr
b&3.31&0.192&76.4&$38.6$&$5.06\times 10^{-4}$&
$5.74 \times 10^{-2}$&0.468\cr
c&0.976&$4.44\times 10^{-2}$&44.4&$518.$&$1.47
\times 10^{-3}$&$3.48
\times 10^{-1}$&0.252\cr
\skip{3}
\lline
}
}$$
\vskip 12pt
$^a$ Parameters for the galaxy models given in the first
column are listed.  From the second to the eighth column, respectively,
these are: the B--band luminosity, the core radius, the total radius, the
central stellar density, the central gas density, the central dark
matter halo density, and the mean temperature of the stellar distribution.
\vskip 6pt
$^b$ All densities are given in units of $10^{-21}~\rm g~cm^{-3}$.
\vfill\eject
\centerline{REFERENCES}
\def \pp{\par        \noindent \hangindent .4in \hangafter 1}
\def \abc#1#2#3#4 {\pp#1 {#2}, {#3} #4}
\vskip .2in
\abc
{Bender, R., Burstein, D., \& Faber, S. M. 1992} {ApJ} {399} 462
\pp
Binney, J., \& Tremaine, S. 1987, Galactic Dynamics (Princeton
University Press), 228
\pp
Bregman, J. N., McNamara, B. R., \& O'Connell, R. W. 1990, ApJ, 351,
406
\pp
Bruzual, A. 1985, Rev. Mexicana Astr. Ap., 10, 55
\pp
Canizares, C. R., Fabbiano, G., \& Trinchieri, G., 1987, ApJ,
312, 503
\pp
Ciotti, L., D'Ercole, A., Pellegrini, S., \& Renzini, A.,
1991, ApJ, 376, 380
\pp
David, L. P., Forman, W., \& Jones, C., 1991, ApJ, 369, 121
\pp
Davies, R. L., Sadler, E. M., \& Peletier, R. F. 1993, MNRAS, 262, 650
\pp
de Jong, T. 1986, The Spectral Evolution of Galaxies, ed. C. Chiosi, \&
A. Renzini, (Dordrecht: Reidel), 111
\pp
de Jong, T., Norgaard--Nielsen, H. U., Jorgensen, H. E., \& Hansen, L.
1990, A\&A, 232, 317
\pp
Djorgovski, S., \& Davis, M., 1987, ApJ, 313, 59
\pp
Donnelly, R. H., Faber, S. M., \& O'Connell, R. M., 1990, ApJ, 354, 52
\pp
Draine, B. T., \& Anderson, N. 1985, ApJ, 292, 494
\pp
Draine, B. T., \& Salpeter, E. E. 1979, ApJ, 231, 77
\pp
Dressler, A., Lynden--Bell, D., Burstein, D., Davies, R.L., Faber, S. M.,
Terlevich, R. J., \& Wegner, G. 1987, ApJ, 313, 42
\pp
Dwek, E. 1986, ApJ, 302, 363
\pp
Dwek, E. 1987, ApJ, 322, 812
\pp
Dwek, E. 1993, private communication
\pp
Dwek, E., \& Werner, M. W. 1981, ApJ, 248, 138
\pp
Ebneter, K., Djorgovski, S., \& Davis, M. 1988, AJ, 95, 422
\pp
Edge, A. C. \& Stewart, G. C. 1991, MNRAS, 252, 428
\pp
Ferguson, H. C., \& Davidsen, A. F. 1993, ApJ, 408, 92
\pp
Forman, W., Jones, C., \& Tucker, W. 1985, ApJ, 293, 102
\pp
Forman, W., Schwarz, J., Jones, C., Liller, W., \& Fabian, A. C. 1979,
ApJ, 234, L27
\pp
Gonzalez, J. J., Faber, S. M., \& Worthey, G. 1994, in preparation
\pp
Forman, W., Jones, C., David, L., Franx, M., Makishima, K.,
\& Ohashi, T. 1993, ApJ, 418, L55
\pp
Jura, M. 1986, ApJ, 306, 483
\pp
Jura, M., Kim, D. W., Knapp, G. R., \& Guhathakurta, P. 1987, ApJ, 312,
L11
\pp
Kley, W., \& Mathews, W. G. 1995, ApJ, in press
\pp
Knapp, G. R., Gunn, J. E., \& Wynn--Williams, C. G. 1992, ApJ, 399, 76
\pp
Kormendy, J., \& Djorgovski, S. 1989, ARA\&A, 27, 235
\pp
Kormendy, J., \& Stauffer, J. 1987, in Structure and Dynamics of
Elliptical Galaxies, IAU Symp. No. 127, ed. T. de Zeeuw (Dordrecht:
Reidel), 405
\pp
Lauer, T. R. 1989 in {\it Dynamics of Dense Stellar Systems}, ed. D. Merritt,
(Cambridge: Cambridge University Press), 3
\pp
Loewenstein, M. \& Mathews, W. G., 1987, ApJ, 319, 614
\pp
Loewenstein, M. \& Mathews, W. G., 1991, ApJ, 373, 445
\pp
Makishima, K. 1994, talk presented at New Horizons of X--ray Astronomy,
Tokyo, Japan
\pp
Mathews, W. G., 1988, AJ, 95, 1047
\pp
Mathews, W. G., 1989, AJ, 97, 42
\pp
Mathews, W. G., 1990, ApJ, 354, 468
\pp
Mathis, J. S., Rumpl, W., \& Nordsieck, K. H. 1977, ApJ, 217, 425
\pp
Mushotzky, R. F., Loewenstein, M., Awaki, H., Makishima, K.,
Matsushita, K., \& Matsumoto, H. 1995, ApJL, in press
\pp
Nulsen, P. E. J., Stewart, G. C., \& Fabian, A. C. 1984, MNRAS, 208, 185
\pp
Raymond, J. C. 1991, private communication
\pp
Renzini, A. \& Buzzoni, A. 1986, The Spectral Evolution of Galaxies, ed.
C. Chiosi, \& A. Renzini, (Dordrecht: Reidel), 135
\pp
Renzini, A. \& Ciotti, L., 1993, ApJL, 416, L49
\pp
Sadler, E. M., \& Gerhard, O. E. 1985, MNRAS, 214, 177
\pp
Sarazin, C. L. 1987, in {\it IAU Symposium 117, Dark Matter in the
Universe}, ed. G. Knapp and J. Kormendy (Dordrecht:Reidel), p. 183
\pp
Seab, C. G. 1987, in Interstellar Processes, ed. D. J. Hollenbach, \& H.
A. Thronson, 491
\pp
Seab, C. G., \& Snow, T. P. 1989, ApJ, 347, 479
\pp
Serlemitsos, P. J., Loewenstein, M., Mushotzky, R. F., Marshall, F. E.
1993, ApJ, 413, 518
\pp
Schombert, J. M., Hanlan, P. C., Barsony, M., \& Rakos, K. D. 1993,
AJ, 384, 433
\pp
Tielens, A. G. G. M., McKee, C. F., Seab, C. G., \& Hollenbach, D. J.
1994, ApJ, in press
\pp
Trinchieri, G., Fabbiano, G., \& Canizares, C. R., 1986, ApJ, 310, 637
\pp
van der Marel, R. P., 1991, MNRAS, 253, 710
\vfill\eject
\centerline{FIGURE CAPTIONS}
\vskip .2in
{\bf Figure 1:} The ($L_B$,$\langle T_* \rangle$)-relation for the model
galaxies of Appendix A is shown.  The filled circles (connected by the
solid line) give results for some of our models.  The open circles
give the observed $T$ and $L_B$ for NGC 4636, NGC 4406, and NGC 4472.

{\bf Figure 2:} The solid line gives the inward velocity of the gas in
the cooling flow of galaxy a (see Table 1) plotted as a function of
the radius normalized to the outer radius.  The dashed and dashed--dotted
lines give the velocities for galaxies b and c, respectively.

{\bf Figure 3:} The sputtering time (defined by eq. [15] of the text)
and flow time (eq. [16]) are shown by dashed and solid lines,
respectively.  The heaviest lines give the results of galaxy a, the second
heaviest lines correspond to galaxy b, and the lightest lines correspond
to galaxy c.

{\bf Figure 4:} The heating rate (plotted versus grain size) due to
absorption of ambient starlight is given as the solid line, the heating
due to collisions with energetic electrons is given by the
dashed--dotted line,
and an upper bound for heating due to absorption of thermal X--rays is
given by the dashed line.  The indicated rates are computed at a radius
of 0.4 times the outer radius of galaxy a (see Table 1).

{\bf Figure 5a:} The grain size distribution (graphite) at various galactic
radii are shown for model galaxy a.
The grain size distribution extends from $a_{min}=
0.001\mu m$ to $a_ {max} = 0.3\mu m$.  The topmost curve gives the number
density of grains per unit grain size at the core radius $r_c$.  The ten lower
curves starting from the top give the grain size distributions at 0.1,
0.2,..., and 1.0 times the outer radius $r_t$.

{\bf Figure 5b:} The grain size distributions (graphite) for model
galaxy c are shown at various radii.  The topmost curve applies at the
core radius, and the ten successively lower curves correspond to 0.1,
0.2,..., and 1.0 times the outer radius.

{\bf Figure 5c:} The grain size distributions (graphite) at 0.5 times
the outer radius for galaxy model a are shown.  The solid line gives
the case where we assume $a_{min}=0.001\mu m$, $a_{max}=0.3\mu m$, and
that the grain size distribution in the stellar ejecta is given by
eq. (39).  The dashed dotted line corresponds to the same case as
the solid line except we assume $a_{max}=1.0\mu m$.  The dashed line
assumes the same power law in the stellar ejecta, but that $a_{min}=
0.08\mu m$ and $a_{max}=1.0\mu m$.  The grain size distribution
in the ISM is, however, computed for smaller grains than $a_{min}$.

{\bf Figure 6:} The rate of energy loss from the hot ISM due to heating
of grains is plotted as a function of radius for galaxy model a.  The
light solid line gives the total heating rate assuming
$a_{min}=0.001\mu m$ and $a_{max}=0.3\mu m$.  The contributions to the
grain heating rate from graphite and silicate grains for this case are
given by the short dashed--dotted and the long dashed--dotted lines,
respectively.  The short dashed line gives the total heating rate in
the case where $a_{min}=0.001\mu m$ and $a_{max}=1.0\mu m$.  The long
dashed line corresponds to taking $a_{min}=0.08\mu m$ and
$a_{max}=1.0\mu m$.  The total heating rates given by the light solid
line, the short dashed line, and long dashed line are nearly
identical.  The rate of energy loss due to X--ray emission
is given by the heavy solid line.
\vfill\eject
\end